\documentstyle[12pt,epsfig]{article}
\def\lsim{\raise0.3ex\hbox{$<$\kern-0.75em\raise-1.1ex\hbox{$\sim$}}}
\def\gsim{\raise0.3ex\hbox{$>$\kern-0.75em\raise-1.1ex\hbox{$\sim$}}}

\def\beq{\begin{equation}}
\def\eeq{\end{equation}}
\def\bea{\begin{eqnarray}}
\def\eea{\end{eqnarray}}
\def\bq{\begin{quote}}
\def\eq{\end{quote}}

\newcommand{\rr}{\mbox{\boldmath $r$}}

\newcommand{\rp}{\mbox{\boldmath $p$}}

\def\gappeq{\mathrel{\rlap {\raise.5ex\hbox{$>$}}
{\lower.5ex\hbox{$\sim$}}}}

\def\lappeq{\mathrel{\rlap{\raise.5ex\hbox{$<$}}
{\lower.5ex\hbox{$\sim$}}}}

\def\Toprel#1\over#2{\mathrel{\mathop{#2}\limits^{#1}}}

\newcommand{\rk}{\mbox{\boldmath $k$}}

\begin{document}
\pagestyle{empty}
\begin{center}
{\bf NUCLEAR CHARM AND BOTTOM PRODUCTION: A COMPARISON AMONG  HIGH ENERGY APPROACHES}
\\
\vspace*{1cm}
 V.P. Gon\c{c}alves $^{1}$, M.V.T. Machado  $^{1,\,2}$\\
\vspace{0.3cm}
{$^{1}$ Instituto de F\'{\i}sica e Matem\'atica,  Universidade
Federal de Pelotas\\
Caixa Postal 354, CEP 96010-090, Pelotas, RS, Brazil\\
$^{2}$ \rm High Energy Physics Phenomenology Group, GFPAE,  IF-UFRGS \\
Caixa Postal 15051, CEP 91501-970, Porto Alegre, RS, Brazil}\\
\vspace*{1cm}
{\bf ABSTRACT}
\end{center}
 
\vspace*{1mm}
\noindent

\vspace*{1cm}
\noindent
\rule[.1in]{16.5cm}{.006in}

\vspace{-3cm}
\setcounter{page}{1}
\pagestyle{plain}
We calculate the nucleon and nuclear  photoproduction  cross sections for heavy quarks within  the $k_{\perp}$-factorization  formalism, considering  the current high energy approaches which include nuclear and saturation effects.  
Our results demonstrate that  a future experimental analysis of this process would allow to constraint the QCD dynamics at high energies.      
\vspace{1.8cm}

\vspace{-1cm}

\section{Introduction}

The electron-proton ($ep$) collider at HERA has opened up a new kinematic regime in the study of the deep structure of the proton and, in general, of hadronic interactions. This regime is characterized by small values of the Bjorken variable $x= Q^2 / s$, where $Q^2$ is the momentum transfer and $\sqrt{s}$ is the center-of-mass energy. The experimental results from HERA have shown a striking rise of the proton structure function $F_2(x,Q^2)$ for values $x < 10^{-2}$, which implies that the cross section increases faster than logarithmically on energy. This high energy behavior in the hard-scattering regime is expected if the underlying dynamics is driven by self-interacting massless vector bosons, the gluons. Thus, the steep rise of $F_2$ certainly confirms one of the basic predictions of perturbative QCD.  As the predictions of the  collinear  and the $k_{\perp}$-factorization approaches, relying on different assumptions,  agree with the measurements of the inclusive quantities, one current open question is:  what   is the  correct  QCD dynamics at high energies?

In the collinear factorization approach \cite{collfact} all partons involved are assumed to be on mass shell, carrying only longitudinal momenta, and their transverse momenta are neglected in the QCD matrix elements. Moreover, the cross sections for the QCD subprocess are usually calculated in the leading order (LO), as well as in the next-to-leading order (NLO). In particular, the cross sections involving incoming hadrons are given, at all orders, by the convolution of intrinsically non-perturbative (but universal) quantities - the parton densities - with perturbatively calculable hard matrix elements, which are process dependent. The conventional gluon distribution $g(x,\mu^2)$, which drives the behavior of the observables at high energies, corresponds to the density of gluons in the proton having a longitudinal momentum fraction $x$ at the factorization scale $\mu$. This distribution satisfies the DGLAP evolution in $\mu^2$ and does not contain information about the transverse momenta $k_{\perp}$ of the gluon. On the other hand, in the large energy (small-$x$) limit, we have that the characteristic scale $\mu$ of the hard subprocess of parton scattering is much less than $\sqrt{s}$, but greater than the $\Lambda_{QCD}$ parameter. In this limit, the effects of the finite transverse momenta of the incoming partons become important, and the factorization must be generalized, implying that the cross sections are now $k_{\perp}$-factorized into an off-shell partonic cross section and a  $k_{\perp}$-unintegrated parton density function ${\cal{F}}(x,k_{\perp})$, characterizing the $k_{\perp}$-factorization  approach \cite{CCH,CE,GLRSS}.  The function $\cal{F}$ is obtained as a solution  of the   evolution equation associated to the dynamics that governs the QCD at high energies.

Recently, several authors have considered the $k_{\perp}$-factorization approach in order to  analyze  some non-inclusive observables and they have obtained a better description of these quantities than  the collinear approach (For a recent review see Ref. \cite{smallx}). However, the current situation is still not satisfactory, due to the large uncertainty   associated to the lack of a complete knowledge of the unintegrated gluon distribution. A search for this distribution in the nucleon has been  subject of active both  theoretical and phenomenological research in recent years. One of the most  promising process to constraint tha quantity is the heavy quark production in $\gamma p$ interactions \cite{semih,Mariotto_Machado,Timneanu_Motyka}. At high energies, the production of open-flavored $Q\overline{Q}$ pairs is described in terms of photon-gluon fusion mechanism, which is expressed in terms of the unintegrated gluon distribution and the parton-level matrix elements. As the behavior of the unintegrated gluon distribution is determined by the underlying dynamics, in Ref. \cite{Mariotto_Machado} the deviations between the results obtained using a distribution derived from the saturation model and the DGLAP evolution equations were analyzed in detail, as well as compared with the predictions of the collinear approach. Those  predictions reasonably describe the current  HERA data, with sizable deviations appearing  basically at higher  energies. These results motivate us  to analyze the heavy quark production in the   context of photonuclear interactions, where the deviations among the approaches should be amplified due to the presence of the nuclear medium. Here, we estimate the cross section for the nuclear photoproduction of heavy quarks, considering the $k_{\perp}$-factorization approach and distinct nuclear unintegrated gluon distributions. In particular, we consider the unintegrated gluon distribution obtained recently in Ref. \cite{armesto}, where the saturation model was extended to  the nuclear case using the Glauber-Gribov formalism. Moreover, we also consider  the derivative of the EKS nuclear gluon distribution \cite{EKS}, which takes into account  the nuclear medium effects (shadowing, antishadowing, EMC and Fermi motion).  Furthermore, we compare our predictions with those obtained from the  collinear factorization approach. In this case, we consider the EKS nuclear gluon distribution as input in our calculations. We also investigate  the possibility of the nuclear gluon distribution to be  modified by the high parton density  effects, as estimated in  Ref. \cite{ayavic}. This procedure allow us to estimate the theoretical uncertainty present in the predictions of the cross sections at high energies.   
 
This paper is organized as follows. In the next section we present  a brief review of the $k_{\perp}$-factorization approach for the heavy quark photoproduction and estimate the cross section for this process in the nucleon and nuclear cases. In particular, we present the proton unintegrated gluon distribution obtained from the saturation model and  the derivative of the collinear gluon distribution. The results for the total cross sections are compared with the HERA experimental measurements. Similarly, we derive the nuclear unintegrated gluon distribution from the  extension of the saturation model to the  nuclear case and the  corresponding quantity  using the EKS nuclear gluon distribution. Moreover,  we present our predictions for the heavy quark nuclear photoproduction cross section in the $k_{\perp}$-factorization approach and compare them  with those ones from the collinear factorization approach. Finally, in Section \ref{conc} we  summarize the main conclusions.

\section{Nuclear heavy quark photoproduction in the $k_{\perp}$ - factorization approach}

Let us start introducing the theoretical prediction for the heavy quark photoproduction in the $k_{\perp}$-factorization approach for the nucleon (proton) case.  The relevant diagrams are considered with the virtualities and polarizations of the initial partons, taking into account the transverse momenta  of the incident partons. The processes are described through the convolution of off-shell matrix elements with the unintegrated parton distribution, ${\cal F}(x,\rk_{\perp})$ (for a recent review, see \cite{smallx}).  The latter can recover the usual parton distributions in the double logarithmic limit  by its integration over the transverse momentum of the $\rk_{\perp}$ exchanged gluon. The gluon longitudinal momentum fraction is related to the c.m.s. energy, $W_{\gamma \,p}$,  in the photoproduction case as $x=4m_{Q}^2/W_{\gamma \,p}^2$.  A sizeable piece  of the NLO and some of the NNLO corrections to the LO contributions on the collinear approach, related to  the  contribution of non-zero transverse momenta of the incident partons, are already included in the LO contribution within the $k_{\perp}$-factorization approach. Moreover, the coefficient functions and the splitting functions giving the collinear parton distributions are supplemented by all-order $\alpha_s\ln (1/x)$ resummation at high energies \cite{CH}. Consequently, in principle, we expect that in the asymptotic regime of large energies the use of the $k_{\perp}$-factorization implies an enhancement of the cross sections in comparison with the predictions obtained in the collinear factorization \cite{CCH}. This is  associated to the opening of the $k_{\perp}$ phase space (away from the collinear region). 

The    cross section for the heavy-quark photoproduction process is expressed as the convolution of the unintegrated gluon function with the off-shell matrix elements, where the LO matrix elements are well known in the literature \cite{smallx,semih,Mariotto_Machado}.  The       expression for the photoproduction total cross section considering the direct component of the photon can be written as \cite{Mariotto_Machado},
\begin{eqnarray}
&\sigma_{tot}^{phot} (W_{\gamma p})& 
= 
  \frac{\alpha_{em}\,e_Q^2}{\pi}\, \int\, dz\,\,d^2 \rp_{1\perp} \, d^2\rk_{\perp} \, \frac{\alpha_s(\mu^2)\,{\cal F}(x,\rk_{\perp}^2; \,\mu^2)}{\rk_{\perp}^2}\nonumber \\
&&\!\!\!\!\!\!\times 
 \left\{ [z^2+ (1-z)^2]\,\left( \frac{\rp_{1\perp}}{D_1} + \frac{(\rk_{\perp}-\rp_{1\perp})}{D_2} \right)^2 +   m_Q^2 \,\left(\frac{1}{D_1} + \frac{1}{D_2}  \right)^2  \right\}\,, \label{eq:11} 
\end{eqnarray}
where $D_1 \equiv \rp_{1\perp}^2 + m_Q^2$ and $D_2 \equiv (\rk_{\perp}-\rp _{1\perp})^2 + m_Q^2$. The transverse momenta of the heavy quark (antiquark) are denoted by $\rp_{1\perp}$ and $\rp_{2\perp}= (\rk_{\perp}-\rp _{1\perp})$, respectively. The heavy quark longitudinal momentum fraction is labeled by $z$. The scale $\mu$ in the strong coupling constant in general is taken to be equal to the gluon virtuality,  in close connection with the BLM scheme \cite{BLM}. On the other hand, in the leading $\ln (1/x)$ approximation, $\alpha_s$ should take a constant value. In our further analysis on heavy quarks  we use the prescription  $\mu^2=\rk_{\perp}^2 + m_{Q}^2$. In Eq. (\ref{eq:11}), the unintegrated gluon function was allowed to depend also on the scale $\mu^2$, since some parametrizations take this scale into account in the computation of that quantity \cite{smallx}.

\begin{figure}[t]
\begin{tabular}{cc}
\psfig{file=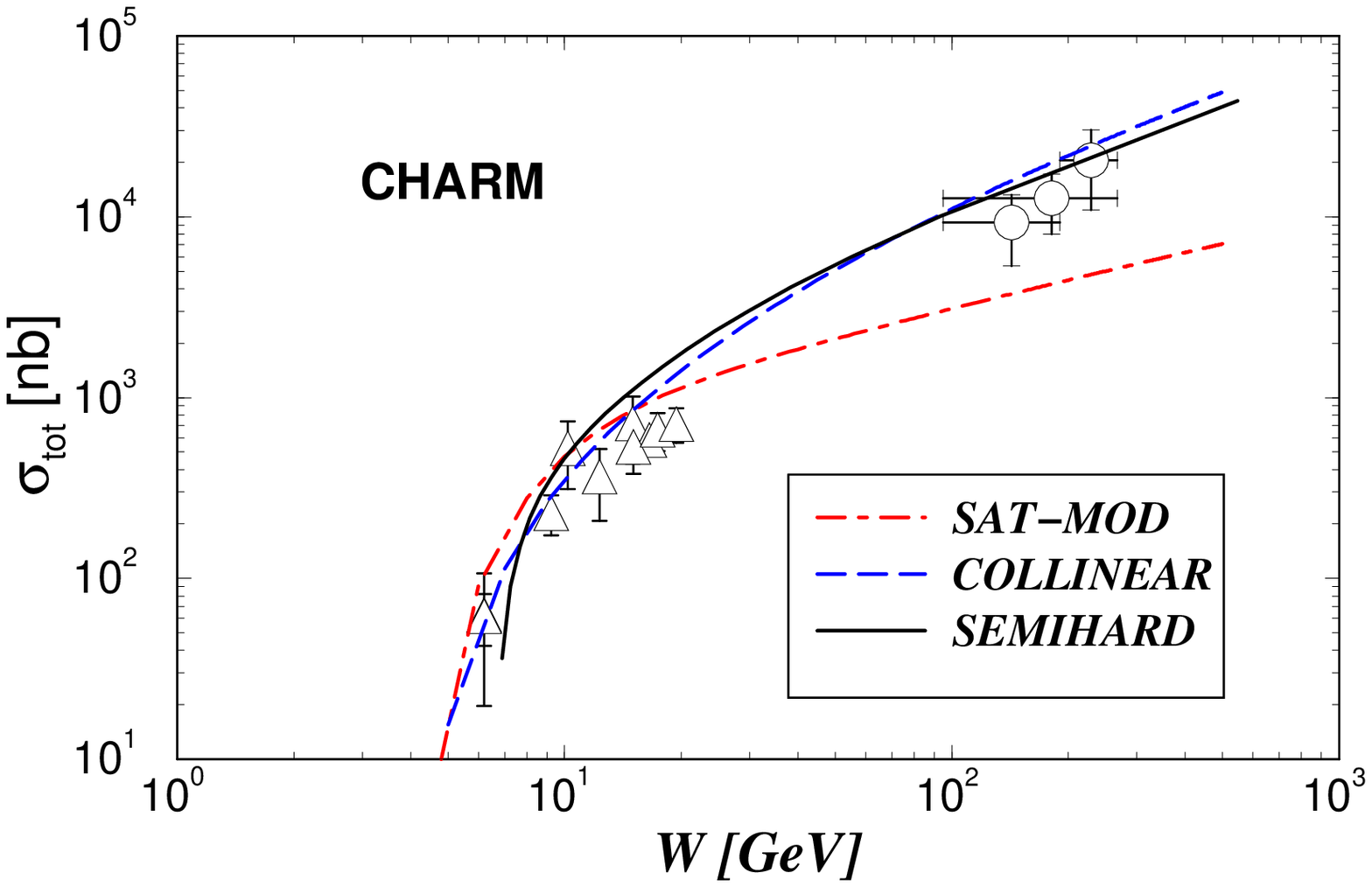,width=80mm} & \psfig{file=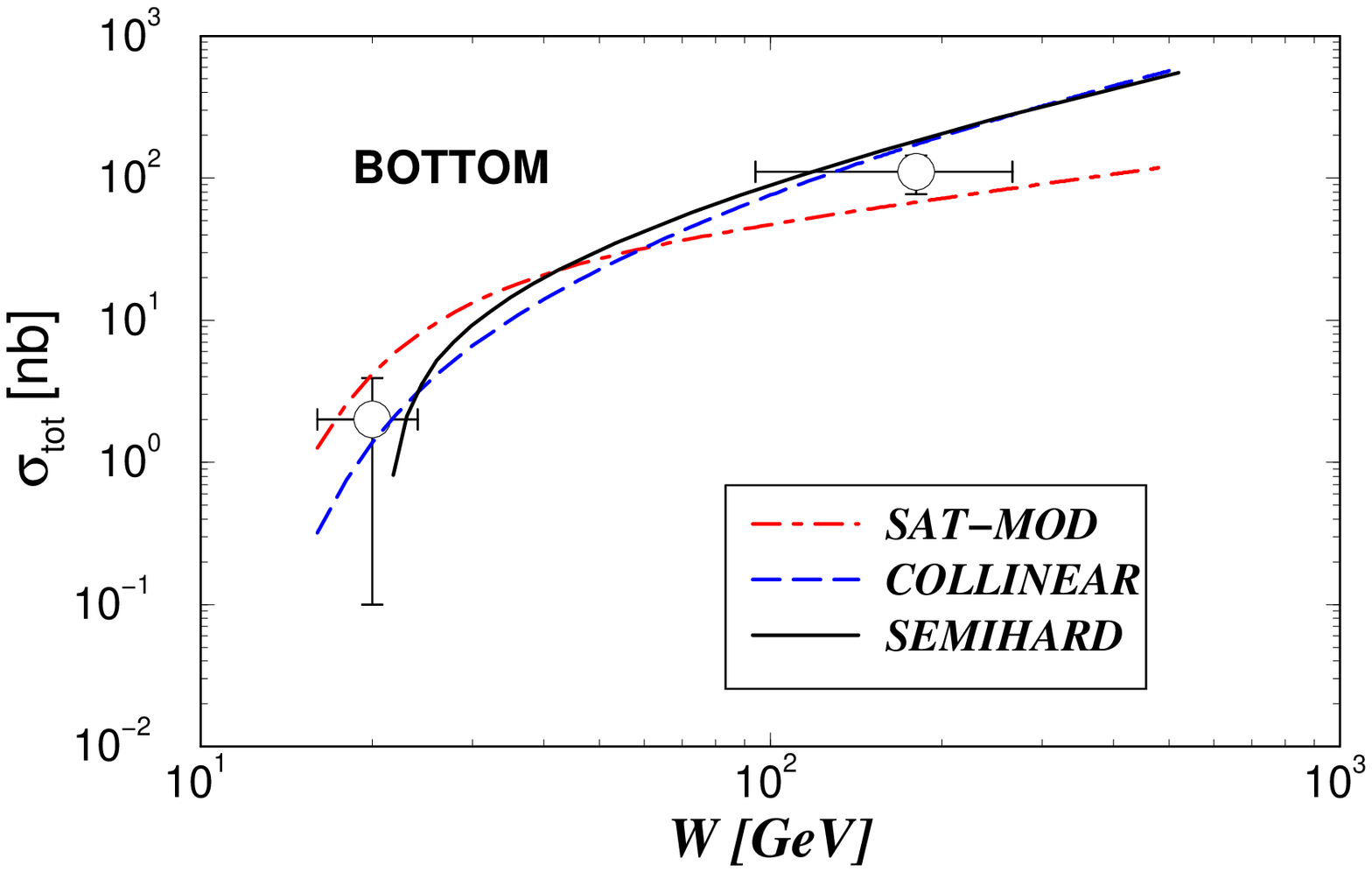,width=80mm}
\end{tabular}
 \caption{\it The charm and bottom photoproduction cross sections as a function of c.m.s. energy $W_{\gamma p}$ in the proton case. In the plots are shown the results for the semihard approach (solid lines), the saturation model (dot-dashed lines) and collinear (long dashed lines). }
\label{fig1}
\end{figure}

In order to perform a phenomenological analysis within the $k_{\perp}$-factorization approach, in the following we use two well known parameterizations for the unintegrated gluon distribution. First, one considers the derivative of the collinear gluon parton distribution function, ${\cal F}_{\mathrm{dgluon}} (x,\rk_{\perp}^2)= \partial\, xG(x,\rk_{\perp}^2)/\partial \ln \rk_{\perp}^2$,   where the GRV94 LO parameterization \cite{GRV94}  was considered for technical simplicity (a comparison using other pdfs can be found in \cite{Mariotto_Machado}). Hereafter, we will denote as semihard the results obtained using this procedure.  It is important to emphasize that if more recent parameterizations are considered, a less steep growth on energy is obtained, but still in agreement with the experimental data \cite{Mariotto_Machado}.  The low transverse momentum region was computed using the ansatz as in Ref. \cite{Mariotto_Machado}. A shortcoming  with this simple parameterization for ${\cal{F}}$   is that it becomes negative  at large  $x$. This can be overcome through the introduction of the doubly logarithmic  Sudakov form factor $T_g(\rk_{\perp}^2, \mu^2)$, which gives the survival probability that the parton with transverse momentum $p_{\perp}$ remains intact in the evolution up to the factorization scale $\mu^2$ \cite{Martin}. So, we can write the unintegrated gluon distribution as  ${\cal F}_{\mathrm{dgluon}} (x,\rk_{\perp}^2)= \partial\,[T_g(\rk_{\perp}^2, \mu^2)\,xG(x,\rk_{\perp}^2)]/\partial \ln \rk_{\perp}^2$. We will denote as semihard the results obtained using this procedure.

  Second, one compute the heavy quark cross sections using the phenomenological saturation model \cite{GBW}, whose unintegrated gluon distribution is given by,
\begin{eqnarray}
 {\cal F}_{\mathrm{sat}}\,(x,\,\rk_{\perp}^2) = \frac{3\,\sigma_0}{4\,\pi^2 \alpha_s} \,\left( \frac{\rk_{\perp}^2}{Q_s^2(x)} \right)\, \exp \left(-\frac{\rk_{\perp}^2}{Q_s^2(x)}  \right) \, (1-x)^7\,,
\label{eq:21}
\end{eqnarray}
where one has  used the parameters from \cite{GBW}, which include the charm quark with mass $m_c=1.5$ GeV. The saturation 
scale $Q_s^2(x)= \left(x_0/x \right)^{\lambda}$ GeV$^2$, gives the onset of the saturation phenomenon to  the process.  The 
last factor in the equation above takes into account the low energy threshold effects.   In Fig. \ref{fig1} the  charm and bottom photoproduction  cross sections \cite{chafix,chahigh,botfix,bothigh}  are shown comparing the referred parameterizations for the unintegrated
 gluon function. As already found in Ref. \cite{Mariotto_Machado}, the saturation model gives a lower bound for the cross sections, 
having  a mild increasing on $W_{\gamma p}$ , whereas the derivative of the usual pdf presents a steeper growth on energy. The latter  
disagree with  the low energy data, but  this can be overcome  by introducing the  Sudakov form factor discussed before. In the plots 
we also present the results considering the collinear approach  for the LO process $\gamma g \rightarrow Q\bar{Q}$.  We have performed 
fully LO calculations, including LO parton densities and a one-loop calculation evaluation of $\alpha_s$. Similar results are obtained using NLO calculations, since the NLO corrections for 
heavy quark photoproduction cross section can be expressed in terms of  $K$- factors, which is about 1.6 for bottom and 
$\sqrt{s} = 1\,$ TeV, while  the NLO parton densities are smaller  by a similar factor than the LO one. Here we have used $m_c=1.5$ GeV, $m_b=4.5$ GeV,  factorization scale $\mu^2_F=\hat{s}$ and the GRV94 LO
 gluon distribution. 
 If the the GRV98 LO  parameterization is used we have that the cross section is reduced by $\approx 20 \%$. In Ref. \cite{victormag3} 
we have analyzed the dependence of the heavy quark cross section  in the  choices of mass, pdfs and factorization scale. 
 We have that the semihard result using ${\cal F}_{\mathrm{dgluon}} (x,\rk_{\perp}^2)$  is quite similar to the collinear one at
 high energies and the deviation with the saturation model is smaller for the bottom case  (For a more detailed discussion see 
Ref. \cite{Mariotto_Machado}). 

Having reviewed the notions of the $k_{\perp}$-factorization approach applied to the heavy quark photoproduction and computed the relevant cross section at the nucleon level, now we will consider an attempt of its extension to the nuclear case. In order to do so, we need a parameterization for the nuclear unintegrated gluon function. Corroborated by the good results using  ${\cal F}_{\mathrm{dgluon}} (x,\rk_{\perp}^2)$ we use the following  ansatz for the nuclear case,
\begin{eqnarray}
 {\cal F}_{\mathrm{nuc}}\,(x,\,\rk_{\perp}^2;\,A) =  \frac{\partial\, xG_{A}(x,\,\rk_{\perp}^2)}{\partial \ln \rk_{\perp}^2}\,,
\label{eq:31}
\end{eqnarray}
where $xG_{A}(x,Q^2)$ is the nuclear gluon distribution, which  was taken from the EKS parameterization \cite{EKS}. Consequently, with this procedure we include in our calculations the medium effects (shadowing, antishadowing, EMC and Fermi motion effects) estimated by this parameterization. Moreover, we emphasize that this nuclear gluon distribution is solution of the DGLAP evolution equations, {\it i. e.} it is associated to a linear dynamics which does not consider dynamical saturation effects.  Similarly to the nucleon case, we denote as semihard the results obtained using this procedure.

In comparison, we calculate the nuclear cross section using the model proposed in Ref. \cite{armesto}, which is an extension of the $ep$  saturation model through  Glauber-Gribov formalism. In this model the cross section for the  heavy quark photoproduction on nuclei targets reads as  \cite{armesto,Goncalves_Machado}
\begin{eqnarray} 
\sigma_{tot}^{\gamma\,A} (W, A)  = \int_0^1
dz\, \int d^2\rr \, |\Psi_{T} (z,\,\rr,\,Q^2=0)|^2 \, \sigma_{dip}^{\mathrm{A}}
(\tilde{x},\,\rr^2,A)\,,
\label{sigmaphot}
\end{eqnarray}
where longitudinal contributions are suppressed and the transverse wave function is known (See e.g. Ref. \cite{predazzi}). The nuclear dipole cross section is given by,
\begin{eqnarray}
\sigma_{dip}^{\mathrm{A}} (\tilde{x}, \,\rr^2, A)  = \int d^2b \,\, 2
\left\{\, 1- \exp \left[-\frac{1}{2}\,A\,T_A(b)\,\sigma_{dip}^{\mathrm{p}} (\tilde{x}, \,\rr^2)  \right] \, \right\}\,,
\label{sigmanuc}
\end{eqnarray}
where $b$ is the impact parameter of the center of the dipole relative to the center of the nucleus and the integrand gives the total dipole-nucleus cross section for fixed impact parameter. The nuclear profile function is labeled by $T_A(b)$, which will be obtained from the 3-parameter Fermi distribution for the nuclear density \cite{Devries}.  The parameterization for the dipole cross section takes the eikonal-like form, $\sigma_{dip}^{\mathrm{p}} (\tilde{x},\,\rr^2)  =  \sigma_0 \,[\, 1- \exp \left(-Q_s^2(\tilde{x})\,\rr^2/4 \right) \,]$, with the saturation scale  previously defined  and  
$ \tilde{x}=(Q^2 + 4\,m_Q^2)/W_{\gamma p}^2$.


The  equation above sums up all the multiple elastic rescattering diagrams of the $q \overline{q}$ pair
and is justified for large coherence length, where the transverse separation $r$ of partons in the multiparton Fock state of the photon becomes as good a conserved quantity as the angular momentum, {\it i. e.} the size of the pair $r$ becomes eigenvalue
of the scattering matrix. It is important to emphasize that for very small values of $x$, other diagrams beyond the multiple Pomeron exchange considered in this formalism should contribute ({\it e.g.} Pomeron loops) and a more general approach for the high density (saturation) regime must be considered. However, we believe that this approach allow us to obtain lower limits of the high density effects in the RHIC and LHC kinematic range. Therefore, at first glance,  the region of applicability of
this  model should be at  small values of  $x$, i.e. large
coherence length, and for not too high  values of virtualities,
where the implementation of the DGLAP evolution in the $ep$ saturation model should be
required. Therefore, the approach is quite suitable for the
analysis of heavy quark photoproduction in the kinematical ranges of the future $eA$  colliders (eRHIC/TESLA). In Ref. \cite{Goncalves_Machado} we have analyzed in detail the behavior of the dipole-nuclei cross section and estimated the nuclear heavy quark photoproduction cross section in the inclusive and diffractive cases. In particular, we have predicted large cross sections at eRHIC/TESLA  energies.

 The corresponding unintegrated gluon distribution can be recovered from a Bessel-Fourier transform to the momentum representation \cite{armesto},
\begin{eqnarray}
 {\cal F}_{\mathrm{nuc}}\,(x,\,\rk_{\perp}^2, b) & = &  -\frac{N_c}{4\pi^2 \alpha_s}\, \rk_{\perp}^2 \, \int \frac{d^2\rr}{2\pi}\,\exp \left(i\, \rk_{\perp}\cdot \rr\right)\,\, \sigma_{dip}^{\mathrm{A}} (\tilde{x}, \,\rr^2, \,A) \nonumber \\
& = &  \frac{N_c}{\pi^2 \alpha_s} \,\left( \frac{\rk_{\perp}^2}{Q_s^2}\right)\, \sum_{m=1}^{\infty} \,\sum_{n=0}^{m}\,\frac{\left(-\frac{1}{2}\,A\,T_A(b)\,\sigma_0\right)^m}{m\, !}\,C^{n}_{m}\,\frac{(-1)^n}{n} \exp\,\left( -\frac{\rk_{\perp}^2}{n\,Q_s^2}\right)\,
\end{eqnarray}
which depends on the transverse momentum $\rk_{\perp}$ through the scaling variable $\tau \equiv  \rk_{\perp}^2/Q_s^2$. The unintegrated gluon vanishes asymptotically at $ \rk_{\perp}^2 \rightarrow 0,\,\infty$ and its maximum can be identified with the saturation scale $Q_{s\,A}(x)$ \cite{armesto,Armesto_Braun}.

 It has been verified in Ref. \cite{Goncalves_Machado}, where one computes heavy quark photoproduction within the saturation approach, that the resummation of high density effects at  the proton level is less sizeable in the final results at nuclear level. Hence, for heavy quark production, we can approximate the dipole nucleon cross section as  $\sigma_{dip}^{\mathrm{p}} (\tilde{x},\,\rr^2)  \simeq  \sigma_0\,Q_s^2(\tilde{x})\,\rr^2/4$ and then  compute analytically the unintegrated gluon distribution.  Following the procedure in Ref. \cite{BGBK} (See also Ref. \cite{pirner}), we obtain that the nuclear unintegrated gluon distribution may be expressed as 
\begin{eqnarray}
 {\cal F}_{\mathrm{nuc}}\,(x,\,\rk_{\perp}^2, b)= \frac{N_c}{2\,\alpha_s\,\pi^2}\,\left(\frac{\rk_{\perp}^2}{Q_{s\,A}^2(x)} \right)\, \exp \left(-\frac{\rk_{\perp}^2}{Q_{s\,A}^2(x)}  \right)\,,
\label{gluonnuc}
\end{eqnarray}
where $Q_{s\,A}^2(x)=\frac{1}{2}A\,T_A(b)\,\sigma_0\,Q_s^2(x)$  define the nuclear saturation scale. This result is consistent with that obtained in Ref. \cite{armesto} in the first scattering approximation. Such an approximation is justified in the heavy quark case, which is dominated by small dipole configurations  (large transverse momentum $\rk_{\perp}^2 \simeq m_Q^2$) (See discussion in Ref. \cite{Goncalves_Machado}). In the case of light quarks, the process receives sizeable contribution from large dipole configurations, i.e. $\rk_{\perp}^2 \leq Q_s^2$, and higher order rescatterings are needed for the proton as rendered by the phenomenological saturation model. It is important to emphasize that the expression (\ref{gluonnuc})  shows  clearly the scaling property on the variable $\tau=\rk_{\perp}^2/Q_{s A}^2$. This implies geometric scaling on $\tau$ in the nuclear heavy quark production, which it already been  shown  in the nucleon case \cite{charmletter}. 

In order to compare our predictions for the nuclear photoproduction cross section obtained using the $k_{\perp}$-factorization approach with the collinear one, let us present a brief review of the latter. In this case the cross section is given by a convolution between the partonic cross section for the subprocess $\gamma g \rightarrow Q \overline{Q}$ and the integrated gluon distribution for the nucleus $xG_A(x,Q^2)$. Here we consider the  EKS and AG  parameterizations for this distribution.   As  discussed before, the EKS parameterization was obtained from a global fit of the nuclear experimental data using the DGLAP evolution equations, which is a linear evolution equation which does not consider dynamical saturation (high density) effects. In Ref. \cite{ayavic} a procedure to include these effects  in the nuclear gluon distribution was proposed, resulting in a paramerization for this distribution (AG parameterization), which also includes those present in the EKS parameterization. The main characteristic of this parameterization is that it predicts a stronger reduction of the growth of the gluon distribution at small values of $x$ than the EKS one. In order to estimate the sensitivity of  nuclear heavy quark photoproduction in these effects, we also use this parameterization as input in our calculations using the collinear factorization approach.

\begin{figure}[t]
\begin{tabular}{cc}
\psfig{file=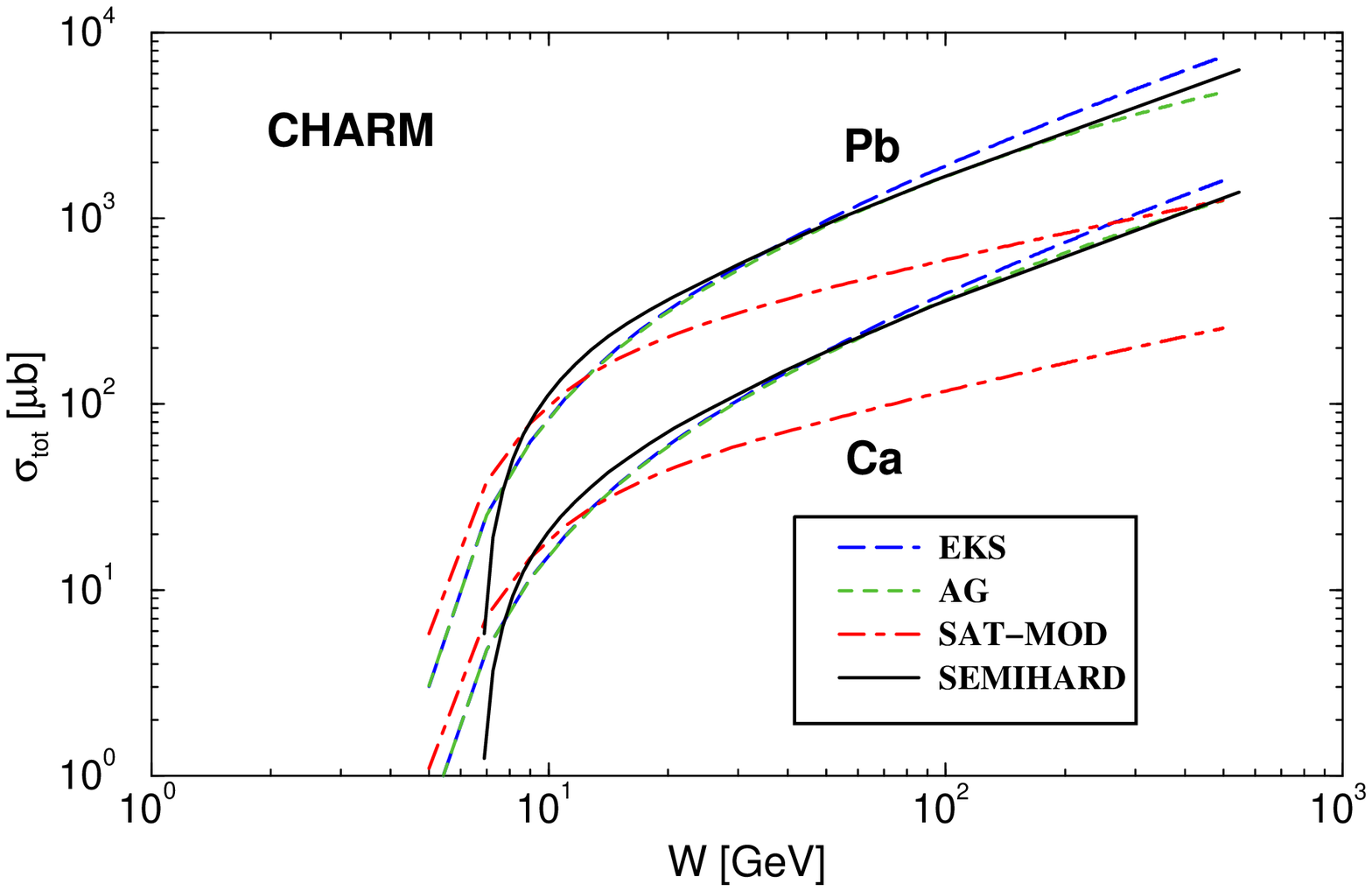,width=80mm} & \psfig{file=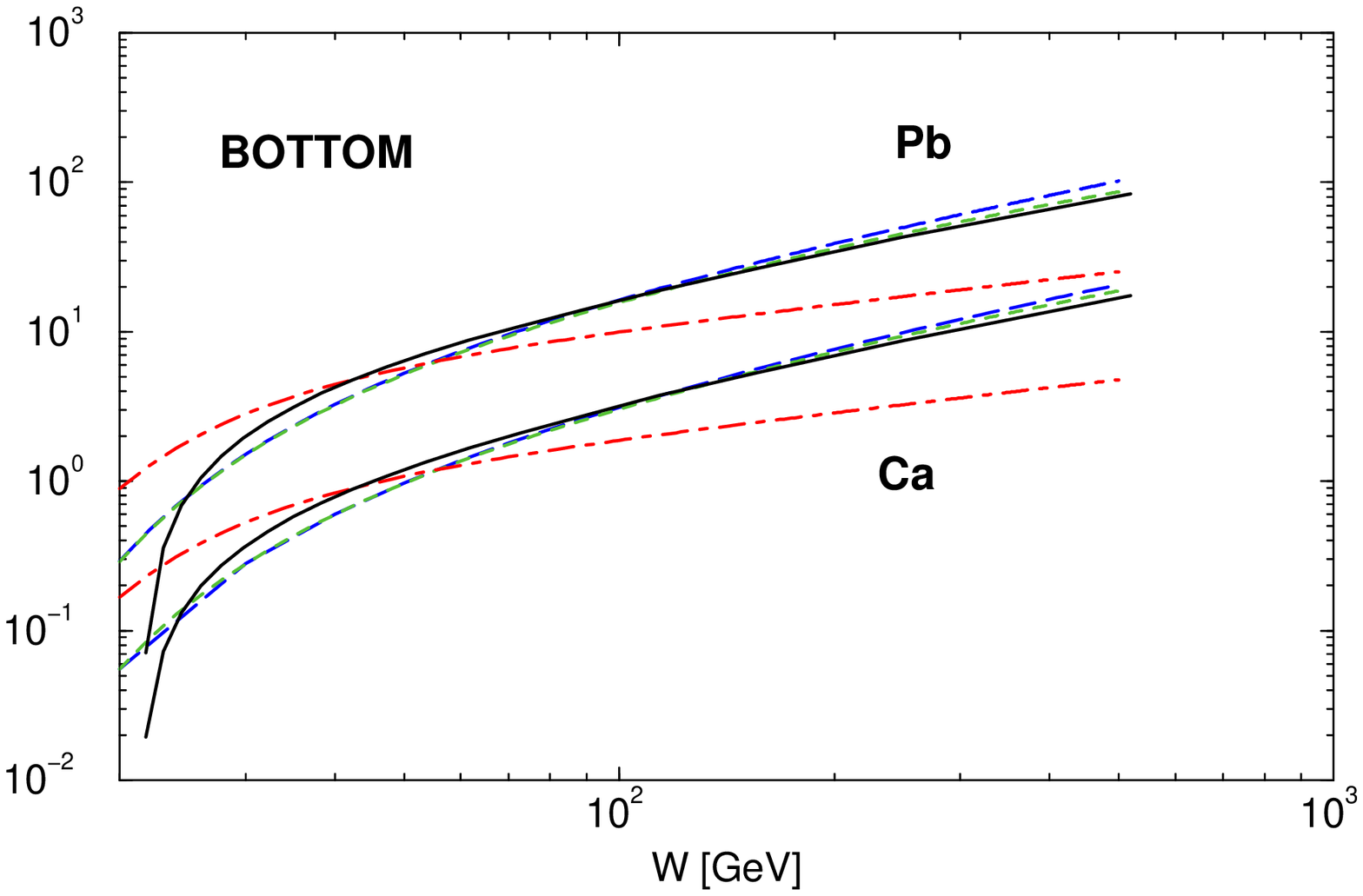,width=77mm}
\end{tabular}
 \caption{\it The charm and bottom nuclear photoproduction cross sections as a function of c.m.s. energy $W$. In the plots are shown the results for the semihard approach (solid lines), the saturation model (dot-dashed lines) and collinear using the parameterizations for the nuclear gluon distribution: EKS (long-dashed lines) and AG (dashed lines). }
\label{fig2}
\end{figure}

In Fig. \ref{fig2} are shown the results for the charm and bottom photoproduction cross section as a function of energy for lead and calcium. We compare the predictions of the  $k_{\perp}$-factorization, considering  the unintegrated gluon distributions discussed above  (SAT-MOD and SEMIHARD in the figure)  with  results of the collinear factorization and two parameterizations for the nuclear gluon distribution (EKS and AG in the figure). 
  We have that the results obtained using the derivative of the EKS gluon distribution and the $k_{\perp}$-factorization (SEMIHARD in the figure)  are
quite similar with those ones coming from the collinear approach where nuclear effects (EKS parameterization \cite{EKS}) and high density corrections (AG parameterization \cite{ayavic}) are taken into account. In particular, we have that the predictions using the AG parameterization in the collinear approach are similar to the semihard one, which does not consider high density effects. This demonstrate that in this process we cannot distinguish if the modification in the behavior of the cross section is associated to high density effects in the collinear approach or a generalization of the factorization without high density effects in the unintegrated gluon distribution. 
On the other hand, if these effects are present and the factorization of the cross section is given by the $k_{\perp}$-factorization, as is the case for the predictions from the saturation model, we have that the difference between the cross sections is large, which should allow to discriminate between the theoretical approaches.  
Therefore, the nuclear cross section would provide a strong test concerning the robustness of the saturation approach in describing the observables. The situation is less clear comparing the semihard  approach and the collinear one. 
One possible interpretation for this result is that the expected enhancement in the semihard approach, associated to the resummation of the $(\alpha_s \, \ln \frac{\sqrt{s}}{m_Q})^n$ in the coefficient function \cite{CCH}, is not sizeable for inclusive quantities  in  the kinematic region of the future colliders. 
Probably, a more promising quantity to clarify this issue would be the transverse momentum $\rp_{\perp}$ distribution. In this case, the semihard approach seems to be in better agreement with experimental data in the $pp$ collisions than the collinear approach \cite{semih}.

\section{Summary and Conclusions}
\label{conc}

We calculate the nucleon and nuclear cross sections for heavy quark photoproduction within the $k_{\perp}$-factorization  approach. Two simple parameterizations for the unintegrated gluon distribution in the proton were considered: the derivative of the collinear gluon pdf and the phenomenological saturation model. The latter underestimate the measured cross section at HERA regime and provides a lower limit for the production. On the other hand, the former agrees with the available high energy data and produces similar results as the collinear approach. In the nuclear case, we have introduced a simple ansatz for the nuclear unintegrated gluon function, namely from the nuclear collinear gluon pdfs. The Glauber-Gribov extension of the saturation model to the nuclear collision is also analyzed, where we found an analytic expression for the unintegrated gluon function using a non-saturated ($\rr^2 \rightarrow 0$) proton dipole cross section. Such a function shows a clear geometric scaling property on the variable $\tau=\rk^2/Q_{s A}^2$. The nuclear cross section for lead and calcium in the energy range relevant for the future $eA$ colliders eRHIC/THERA were computed, comparing them with the results from the collinear approach which take into account nuclear medium effects and also high parton density phenomenon. Our results indicate that   a future experimental analysis of this process can be useful  to constraint the QCD dynamics at high energies.   

\section*{Acknowledgments}
 M.V.T.M. thanks the support of the High Energy Physics Phenomenology Group at the Institute of Physics, GFPAE IF-UFRGS, Porto Alegre. This work was partially financed by the Brazilian funding agencies CNPq and FAPERGS.

\end{document}